\documentclass[twocolumn,aps,pra,superscriptaddress,amsfonts,longbibliography]{revtex4-1}

\usepackage{amsmath}
\usepackage{amssymb}
\usepackage{graphicx}
\usepackage{bm}
\usepackage{array}
\usepackage{dcolumn}
\usepackage{hepparticles}
\usepackage{heppennames}
\usepackage{hepnicenames}
\usepackage{epstopdf}
\usepackage{xcolor}

\usepackage{svg}
\usepackage[T1]{fontenc}
\usepackage{lmodern}

\newcommand{\ket}[1]{{| {#1} \rangle}}

\begin{document}

\title{In-Situ Rewiring of Two-Dimensional Ion Lattice Interactions Using Metastable State Shelving}

\author{Ilyoung Jung}
\affiliation{Indiana University Department of Physics, Bloomington, Indiana 47405, USA}
\author{Antonis Kyprianidis}
\affiliation{University of Louisville Department of Physics and Astronomy, Louisville, Kentucky 40208, USA}
\author{Frank G. Schroer}
\affiliation{Indiana University Department of Physics, Bloomington, Indiana 47405, USA}
\author{Thomas W. Burkle}
\affiliation{Indiana University Department of Physics, Bloomington, Indiana 47405, USA}
\author{Jack Lyons}
\affiliation{Indiana University Department of Physics, Bloomington, Indiana 47405, USA}
\author{Philip Richerme}
\affiliation{Indiana University Department of Physics, Bloomington, Indiana 47405, USA}
\affiliation{Indiana University Quantum Science and Engineering Center, Bloomington, Indiana 47405, USA}

\date{\today}

\begin{abstract}
Trapped-ion lattice geometries, which determine the interactions between trapped-ion qubits, are typically governed by the balance of Coulomb repulsion forces with the external trapping potential. Here we demonstrate how the effective ion lattice geometry and resulting qubit-qubit interactions may be reconfigured in-situ, by shelving specific ions in metastable states outside the qubit subspace. Using a triangular lattice of three $^{171}$Yb$^{+}$ ions, we optically pump selected ions into the long-lived $^2F_{7/2}$ state. We then apply a global Ising-like Hamiltonian to the system and verify that the shelved qubits are fully removed from participation in the quantum dynamics. We characterize the metastable state lifetime in the presence of laser-driven ion-ion interactions, finding a deshelving rate that is orders of magnitude slower than the spin-spin interaction rate and scales quadratically with applied laser intensity.

\end{abstract}

\maketitle

\label{sec:intro}
\section{Introduction}

The observable properties of quantum spin systems are strongly dependent on their underlying lattice geometry \cite{diep2013frustrated}. For instance, anti-ferromagnetic (AFM) Ising interactions on a square lattice lead to simple N\'eel ordering, while on a triangular lattice they lead to frustration and highly entangled ground states \cite{moessner2001ising}. 
On a Kagom\'e lattice, Heisenberg AFM interactions are believed to produce a type of translationally invariant paramagnetic ground state, the so-called quantum spin liquid \cite{balents2010spin,huerga2016staircase}. However, detailed solutions to such correlated many-body systems are often elusive since they may require classical computational resources that grow exponentially with system size. For this reason, a broad range of quantum simulator devices has been developed to study quantum spin physics with differing interactions and dimensionalities \cite{georgescu2014quantum}. In particular, Rydberg-atom systems have shown great flexibility in reconfiguring lattice spin geometries by rearranging atom positions using steerable optical tweezers \cite{ebadi2021quantum,scholl2021quantum}.

Trapped-ion quantum simulators have likewise demonstrated promise in simulating the behavior of interacting spin systems \cite{monroe2021programmable}. In most cases, the ions are trapped in one-dimensional (1D) chains or two-dimensional (2D) triangular arrays, which are the self-assembled lowest-energy configurations of a global confining potential. When driven with global entangling operations, the resulting spin-spin connection graph typically mirrors the ions' physical arrangement. This approach has enabled many experiments probing long-range Ising, XY, and Heisenberg-type spin models, in both 1D chains \cite{monroe2021programmable,BPLayon2011UniDigiQsim,islam2013emergence,richerme2014non,jurcevic2014quasiparticle} and 2D triangular lattices \cite{britton2012engineered,qiao2024tunable,Guo2024TwoDQsim}.

To date, it has proven challenging for trapped-ion systems to explore spin-spin interaction geometries which deviate substantially from the underlying Coulomb crystal configuration. An early example, limited to a 1D chain of three ions, used globally-driven M\o lmer-S\o rensen interactions \cite{molmer1999multiparticle} with a carefully chosen laser detuning to mimic a 2D triangular graph \cite{kim2010quantum}. More recently, theoretical and experimental efforts have proposed global multi-frequency M\o lmer-S\o rensen drives \cite{shapira2018robust,shapira2020theory, Antonis2024EngineerIntgraph,wu2025qubits,shapira2025programmable} and hybrid digital-analog protocols \cite{parra2020digital,bassler2023synthesis,richerme2025multi,solomons2025full} to engineer arbitrary Ising-like couplings, at the expense of increased hardware complexity and experimental runtime. In the most general case, a universal quantum computer may be used to apply a Trotterized sequence of all desired two-body lattice interactions. However, this approach is the most resource intensive and requires $\mathcal{O}(N^2)$ high-fidelity local entangling operations per iteration to replicate arbitrary interaction graphs.

\begin{figure}[t]
    \centering
    \includegraphics[width=\columnwidth]{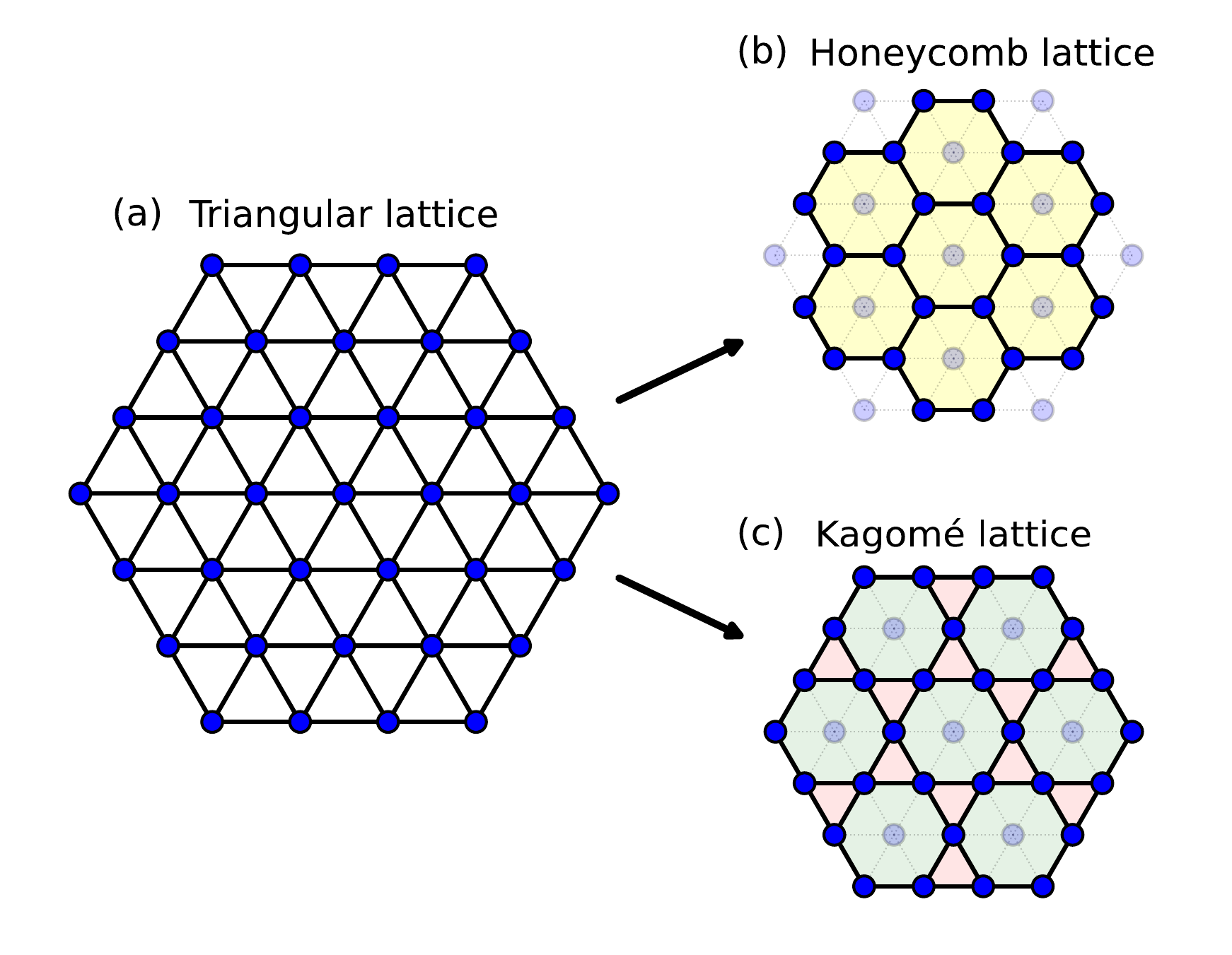}
    \caption{Concept of lattice rewiring via selective ion shelving. (a) Near the center of a large 2D Coulomb crystal, ions self-assemble into a triangular lattice geometry. Each lattice site encodes an effective spin-1/2 system (blue circles). (b)--(c) When an ion is shelved outside the spin-1/2 subspace (gray circles), it no longer interacts with the remaining effective spins. Shelving specific ions within a triangular array provides access to (b) honeycomb and (c) Kagom\'e lattice geometries, among others.}
    \label{fig:concept}
\end{figure}

Here we demonstrate a direct approach to modifying the spin-spin interaction geometry of a trapped-ion crystal, by shelving unwanted ions to a long-lived metastable state outside of the qubit subspace.  Historically, such metastable state shelving has been considered for omg- and dual-type qubit architectures \cite{allcock2021omg,yang2022realizing}, high-fidelity state readout \cite{edmunds2021scalable}, detection of subspace leakage errors \cite{chen2025randomized}, and potentially the simulation of open and disordered quantum systems. In our application, proposed theoretically in \cite{Bermudez2012spinladder,richerme2016two}, we create an effective rewiring of spin-spin interactions without requiring coherent quantum operations or physical rearrangements of ion positions. For example, 
honeycomb or Kagom\'e lattice geometries with fully tunable power-law interactions \cite{monroe2021programmable,Antonis2024EngineerIntgraph,shapira2025programmable} may be realized using an underlying 2D triangular ion array (see Fig.~\ref{fig:concept}).

We implement metastable state shelving of $^{171}$Yb$^{+}$ ions by optically pumping them into the long-lived $^2F_{7/2}$ manifold. Following our shelving protocol, all ions are irradiated with global laser beams to drive Ising-like interactions between the effective spins that remain in the qubit subspace. The resulting system dynamics reveal that shelved ions are fully removed from participation in the evolution, while unshelved ions retain their original couplings under the global drive and evolve as predicted under the applied Hamiltonian. We also verify that in the presence of the global Hamiltonian drive, the shelved ion lifetime remains two to three orders of magnitude longer than the typical spin-spin interaction time over a wide range of applied laser intensities.

This paper is organized as follows. In section II, we introduce our experimental setup, including the ion crystal geometry, state preparation and measurement, and implementation of metastable state shelving. In section III, we describe the implementation of Ising Hamiltonians and investigate the spin dynamics of two- and three-ion systems in the presence of metastable state shelving. In section IV, we characterize the shelved ion lifetime as a function of applied global drive intensity. Finally, section V summarizes our results and outlines considerations for scaling to larger experiments.

\section{Experimental Setup}
\label{sec:ExperimentalSetup}
Experiments are performed using two- and three-ion crystals of $^{171}$Yb$^{+}$ confined in an open-endcap rf blade trap \cite{Xie2021open-endcap}, with secular frequencies \{$\omega_x, \omega_y, \omega_z$\} = 2$\pi \times \{0.978, 1.748, 1.798\}$ MHz. In this configuration, ions form a linear chain along the trap axis (x-axis) when $N=2$ and a lateral 2D crystal when $N=3$. Effective quantum spins are encoded in the hyperfine `clock' states $^2S_{1/2}\ket{F = 0, m_F = 0} \equiv \ket{\downarrow}_{z}$ and $^2S_{1/2}\ket{F = 1, m_F = 0} \equiv \ket{\uparrow}_{z}$ with a frequency splitting $2\pi \times 12.6~\mathrm{GHz}$. A magnetic field $B_0 = 3.6$~G is oriented perpendicularly to the optical table to lift the degeneracy of the $F=1$ hyperfine states and define the quantization axis. Doppler cooling, optical pumping, and spin-state fluorescence detection are all performed using laser light near 369 nm \cite{olmschenk2007manipulation}.

Single- and two-qubit gates are implemented using two-photon Raman transitions with a mode-locked laser at 355 nm \cite{campbell2010ultrafast}. The laser output is split into two non-copropagating Raman beams, each with a beam waist of 30 $\mu$m and power of 80 mW. When the frequency beatnote between Raman beams is tuned to the qubit splitting, we measure a two-photon Rabi frequency of \mbox{$\Omega= 2\pi \times$ 76 kHz}.

\begin{figure}[t]
    \centering
    \includegraphics[width=\columnwidth]{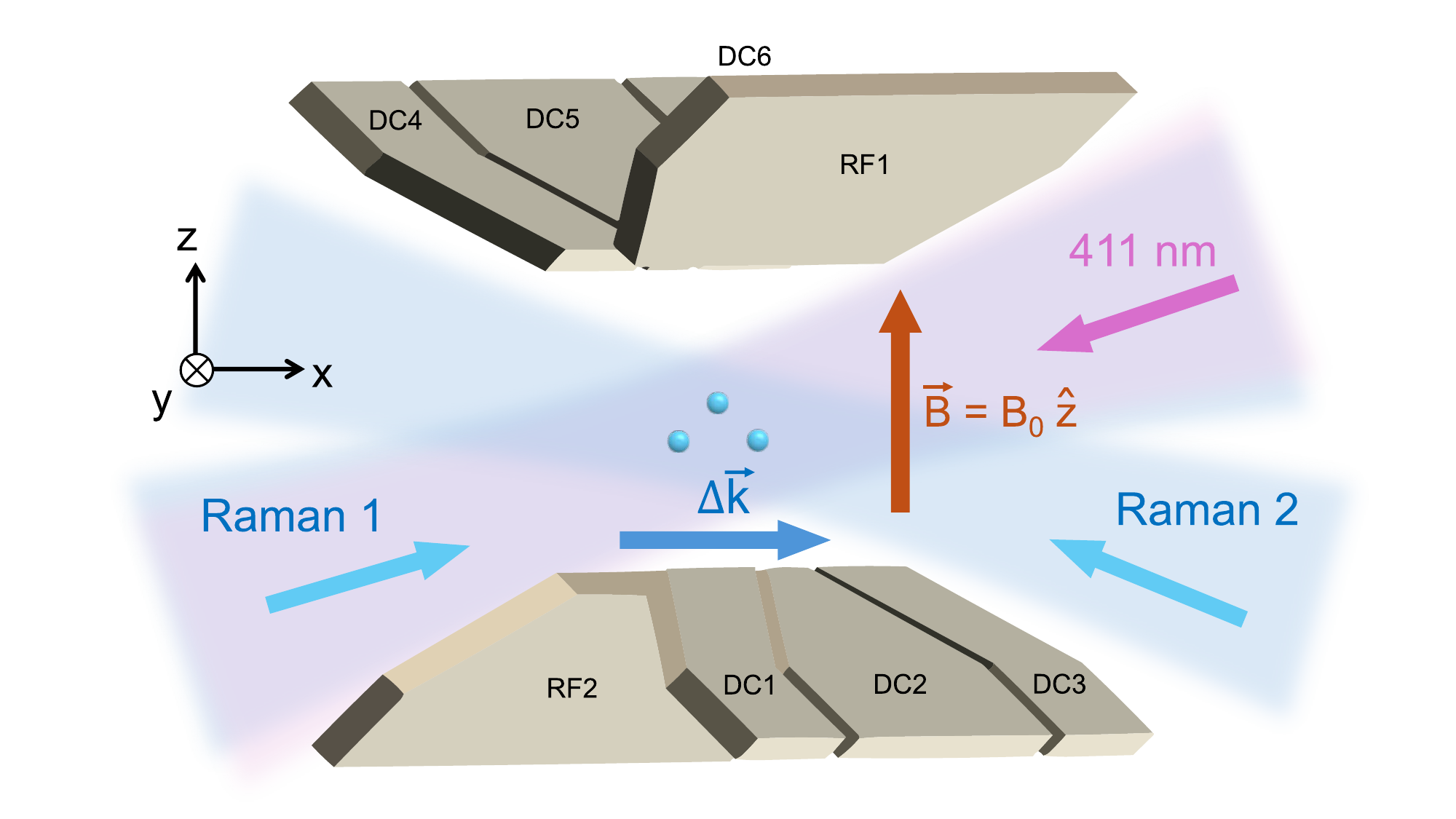}
    \caption{Experimental setup geometry for a three-ion triangular crystal. All beams propagate in and are linearly polarized along the $xy$-plane (no $\pi$ component). The ion crystal lies on the $xz$-plane. Non-copropagating Raman beams have a wavevector difference $\Delta\vec{k}$ along the $\hat{x}$ direction and couple to the axial crystal modes.}
    \label{fig:trapsketch}
\end{figure}

Effective spin–spin interactions are mediated by the axial vibrational modes, since the wavevector difference between Raman beams points along the trap axis in our setup (Fig.~\ref{fig:trapsketch}). We drive global M\o lmer-S\o rensen interactions in the adiabatic regime \cite{Kim2009}, where the effective spin dynamics are well described by the fully-connected Ising Hamiltonian:
\begin{equation}
    \label{eq:ising}
H = \sum_{i<j} J_{ij} \sigma_i^x \sigma_j^x \quad
\text{;} \quad
J_{ij} = \Omega^2R\sum_k \frac{b_{i,k} b_{j,k}}{\mu^2 - \omega_k^2}
\end{equation}
In  Eq.~\ref{eq:ising}, $J_{ij}$ is the effective spin–spin coupling between ions $i$ and $j$, $b_{i,k}$ is the normal mode eigenvector component for ion $i$ and mode $k$, $R=\hbar^2(\Delta \vec{k})^2/(2m)$ is the recoil frequency, $m$ is the ion mass, and $\omega_k$ is the normal mode frequency. The parameter $\mu$ is the laser detuning from the carrier transition, which we choose to couple most strongly to the axial center-of-mass (COM) mode. Under these conditions, the system exhibits nearly uniform all-to-all interactions across all ion pairs, with small deviations arising from residual couplings to higher-frequency axial modes.

\begin{figure}[t]
    \centering
    \includegraphics[width=\columnwidth]{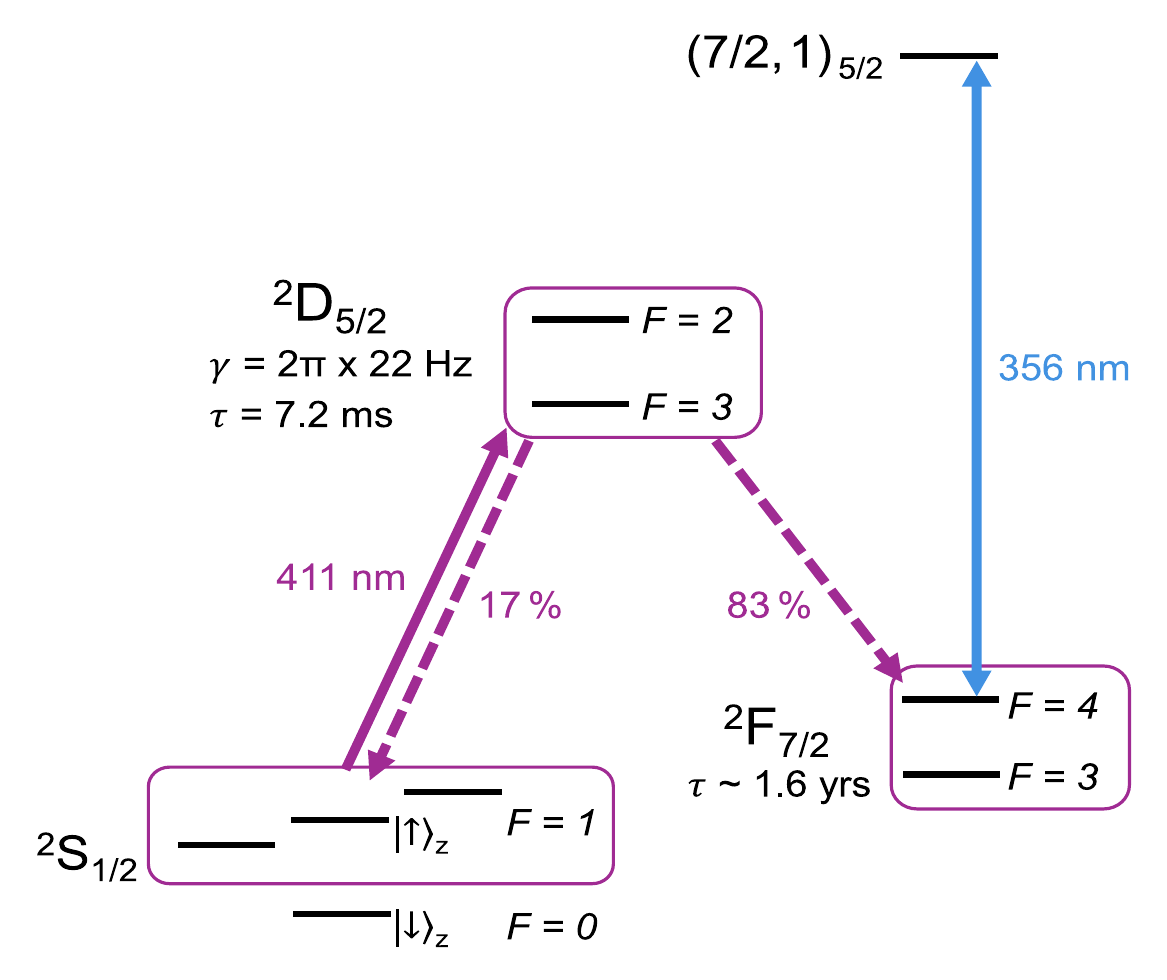}
    \caption{Partial energy level diagram of $^{171}$Yb$^{+}$, showing transitions relevant for optical pumping into the $^2F_{7/2}$ state. The 356~nm transition is relevant to deshelving out of the $^2F_{7/2}$ state and is discussed separately in Sec.~\ref{sec:deshelving}.}
    \label{fig:leveldiagram}
\end{figure}

When we seek to remove qubits from the interaction dynamics, we shelve them into the $^2F_{7/2}$ manifold. Since decay back to the ground state is a strongly forbidden electric octupole (E3) transition, the estimated metastable state lifetime is in excess of 1 year \cite{roberts2000observation,lange2021lifetime}. Rather than directly shelving via the octupole transition, we use a 411 nm laser to optically pump ions through the $^2D_{5/2}$ manifold, which spontaneously decays to $^2F_{7/2}$ with 83\% probability (Figure~\ref{fig:leveldiagram}). High accuracy spectroscopy and branching ratio measurements of this transition have been performed in \cite{tan2021precision}, which we repeat to verify the specific transition frequencies in our experimental system (Appendix~\ref{sec:appx}).

\begin{figure}[t]
    \centering
    \includegraphics[width=0.8\columnwidth, height=5cm]{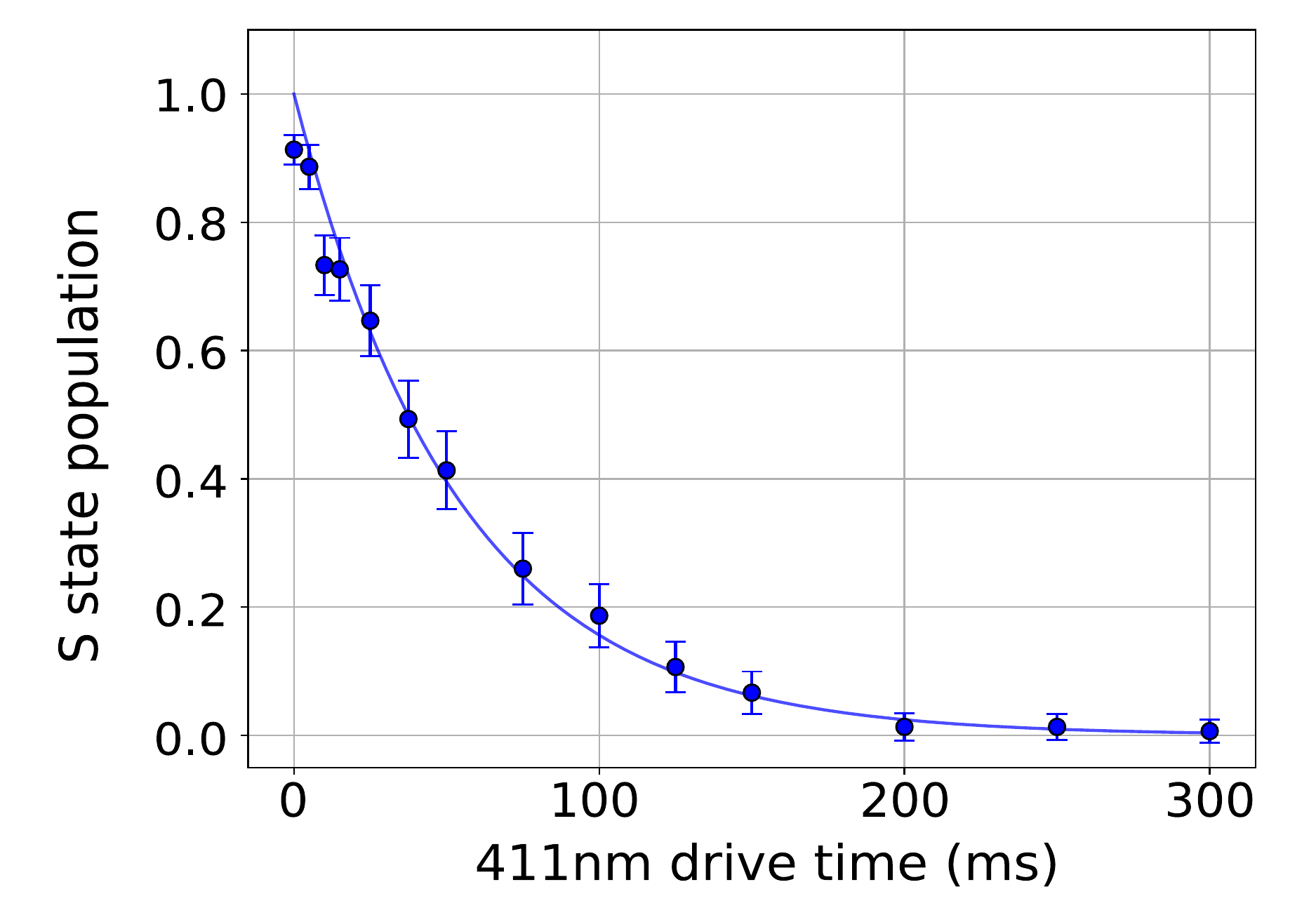}
    \caption{Depopulation of the $^2S_{1/2}$ manifold during optical pumping to the $^2F_{7/2}$ state. The characteristic shelving time is approximately $55~$ms.}
    \label{fig:op}
\end{figure}

Near-unit population transfer into the $^2F_{7/2}$ state is achieved by leaving our $^2S_{1/2}\rightarrow ^2$$D_{5/2}$ optical pumping beam on for much longer than the 7.2~ms $^2D_{5/2}$ state lifetime. For optical pumping, we focus 1.2 mW of 411 nm light to a 30 $\mu$m waist and drive the $\Delta m_F=-1$ transition between $^2S_{1/2}\,\ket{F = 1}\rightarrow$ $^2D_{5/2}\,\ket{F = 3}$. We observe exponential decay of the population in the $S$ manifold with a characteristic time constant of approximately 55 ms (Fig.~\ref{fig:op}). This optical pumping rate is inherently limited by the relatively long lifetime of the $^2D _{5/2}$ state and shows only a weak dependence on applied laser power.

\begin{figure*}[t]
    \centering
    \includegraphics[width=\textwidth]{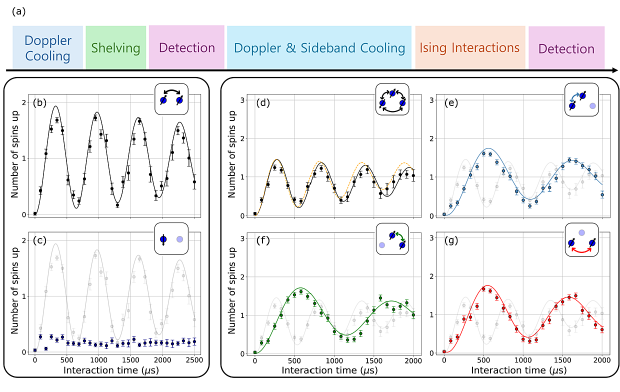}
    \caption{(a) Pulse sequence used for preparing different configurations of shelved ions and measuring their resulting dynamics. The first detection block is used to verify the initial shelving configuration before driving quantum state evolution. (b)--(c) Dynamics of two trapped ions under an applied Ising interaction, without (b) and with (c) metastable state shelving. The gray line in (c) shows the unshelved data for reference. (d)--(g) Dynamics of three trapped ions under an applied Ising interaction. The coherent oscillations differ depending on which ions are shelved. In panel (d), we show the predicted dynamics based on the calculated $J_{ij}$ spin-spin couplings for our trapping parameters (black, solid), as well as the measured $J_{ij}$ couplings extracted from shelving (orange, dashed).}
    \label{fig:withshelving}
\end{figure*}

\section{Dynamics under Metastable State Shelving}
Shelving unwanted ions into long-lived metastable states provides a conceptually simple and intuitive means of spin–spin interaction engineering. Once ions are shelved, the interactions involving the shelved ions are suppressed, thereby modifying the overall interaction graph. Through selective shelving, effectively arbitrary interaction geometries can be realized within a fixed underlying Wigner crystal structure.

In our experiments, we study how metastable state shelving modifies the quantum spin dynamics of small ion crystals. Following Doppler cooling for 5 ms, we apply the 411 nm shelving beam to optically pump ions into the $^2F_{7/2}$ state (Fig.~\ref{fig:withshelving}(a)). Since the 411 nm beam addresses globally in our current setup, we employ a probabilistic strategy for shelving individual ions. First, the optical pumping beam is applied for a limited duration, such that any individual ion has a 30--50\% probability of being shelved. Next, we perform state-dependent fluorescence detection to determine which ions remain in the $^2S_{1/2}$ state. This step identifies which specific ions (if any) were shelved by the 411 nm beam, allowing for post-selection of desired configurations during classical data analysis.

Once the initial configuration of the ion crystal is determined, the ions are re-cooled to near their motional ground state using Doppler and resolved sideband cooling. For ions remaining in the $^2S_{1/2}$ manifold, sideband cooling also initializes them to the state $\ket{\downarrow}_{z}$. Quantum dynamics are then induced by applying a global M{\o}lmer-S{\o}rensen drive to yield effective Ising interactions (Eq.~\ref{eq:ising}) between unshelved ions. Following time evolution of the quantum state, we detect the fluorescence from the ion crystal to determine its final spin projection along the $z-$direction and to verify that the initial shelving configuration remains intact. The data at each timestep is averaged over 150 repetitions (two-ion experiments) or 120 repetitions (three-ion experiments), with state preparation and measurement (SPAM) errors of 4\% and laser intensity noise of $\approx 1$\% contributing to the overall uncertainty.

We begin with the simplest possible system and investigate the quantum dynamics of a two-ion chain, with and without metastable state shelving. In Fig.~\ref{fig:withshelving}(b), when no shelving is applied, we observe coherent spin dynamics between the $\ket{\downarrow\downarrow}_{z}\rightarrow\ket{\uparrow\uparrow}_{z}$ states and a measured Ising coupling strength of $J_{12} = 2\pi\times 0.75$~kHz. The solid black line shows the predicted Ising dynamics, including decoherence effects with a time constant of 5.5 ms. In contrast, when either of the ions is shelved into the metastable state, the system is reduced to an effective single spin with no available two-body interactions (Fig.~\ref{fig:withshelving}(c)). As a result, we observe no coherent oscillations, apart from a small residual offset attributed to spin-motion coupling errors during the adiabatic dynamics.

We then extend our study to a triangular lattice of three ions. When no shelving beam is applied (Fig.~\ref{fig:withshelving}(d)), we again observe coherent quantum dynamics of all ions under a three-particle Ising Hamiltonian (Eq.~\ref{eq:ising}). The solid black curve shows the predicted dynamics based on our trapping configuration, where we calculate all ions to be coupled with approximately equal strengths of $J_{ij}\approx2\pi\times0.45$~kHz. When we apply our probabilistic shelving protocol to this triangular crystal, the most frequent result is for one of the three ions to be pumped into the $^2F_{7/2}$ state. In figures \ref{fig:withshelving}(e)--(g), we show the observed quantum dynamics when one of the three ions has been shelved. In each case, what remains are coherent two-ion dynamics that reveal the underlying spin-spin couplings $\{J_{12},J_{13},J_{23}\}=2\pi\times\{0.47, 0.44, 0.49\}$ kHz. When these directly measured coupling strengths are inserted back into the Hamiltonian of Eq.~\ref{eq:ising}, the resulting dynamics (orange dashed line in Fig.~\ref{fig:withshelving}(d)) show close agreement with the three-spin dynamics data as well as the first-principles theory prediction (solid black line). We additionally confirm that when two or more ions are shelved during our probabilistic protocol, all spin-spin interaction dynamics are suppressed (just as in Fig.~\ref{fig:withshelving}(c)).

Beyond reconfiguring the lattice interactions in-situ, the data in Figs.~\ref{fig:withshelving}(e)--(g) show how metastable state shelving enables direct measurement of individual spin-spin coupling strengths within an ion crystal. Earlier works, in both optical and hyperfine qubits, have characterized the Ising interactions $J_{ij}$ by shelving all ions except $i$ and $j$ into auxiliary Zeeman levels using individually addressed, coherent $\pi$-pulses \cite{jurcevic2014quasiparticle,smith2016many}. By replacing the coherent pulses with an optical pumping beam, we dissipatively prepare the shelved state while avoiding potential sensitivities to laser intensity or timing errors.

For larger systems, it will become crucial to implement our metastable state shelving protocol using individually addressed beams. This is because the success probability of randomly shelving $m$ specific ions from an $N$-ion lattice becomes negligibly small for even moderately sized systems. In contrast, deterministic creation of arbitrary lattices with near unit probability may be accomplished by focusing the 411 nm beam onto specific ions. We estimate that a $\sim 1~\mu$m diffraction-limited beam waist is achievable using our NA=0.28 imaging objective \cite{lee2016engineering}. Moreover, multiple sites may be addressed in parallel by passing the shelving beam through a commercially available acousto-optic deflector or digital micromirror device, for which intensity crosstalk of $<10^{-4}$ has been demonstrated \cite{shih2021reprogrammable}. If further crosstalk suppression is required, a tightly focused 355 nm beam could selectively prepare the target ions in $\ket{1}$ before 411 nm shelving, reducing crosstalk to negligible levels even for systems with hundreds of ions. In this configuration, the intensity of each 411 nm individual addressing beam would be similar to that of the global beam used in this work; as a result, the total shelving time for a large lattice is expected to remain comparable to the 55 ms demonstrated above.

\section{Metastable State Deshelving}
\label{sec:deshelving}
To reconfigure a spin-spin interaction graph using metastable state shelving, it is imperative that the hidden qubits remain shelved throughout the quantum dynamics. If shelved ions return to the ground state manifold prematurely, they re-enter the qubit subspace during the interaction and unpredictably distort the targeted interaction graph. Since all metastable levels will eventually decay to the ground state, candidate metastable states should have natural lifetimes that are long compared to the typical millisecond timescales of quantum simulation experiments (see Fig.~\ref{fig:withshelving} and Ref. \cite{monroe2021programmable}). For example, this consideration strongly disfavors the use of the $^2D_{5/2}$ state ($\tau=7.2$~ms) for metastable state shelving, compared to our chosen $^2F_{7/2}$ state ($\tau=1.6$~years \cite{lange2021lifetime}).

Additional mechanisms beyond simple spontaneous decay may also result in the return of a shelved ion back into the qubit subspace. For instance, the high-power 355 nm Raman beams applied during a quantum simulation may off-resonantly excite shelved ions to a new state. One candidate for this mechanism is the $(7/2,1)_{5/2}$ state, which lies 356 nm above the $^2F_{7/2}$ manifold \cite{Biemont1998} (see Fig.~\ref{fig:spectroscopy}(a)). If excited, ions may undergo a series of decays to the ground state much faster than the natural octupole decay rate, potentially imposing a timescale constraint for quantum simulation experiments. It is therefore critical to confirm that the deshelving timescale in the presence of driven quantum evolution remains long compared to the quantum evolution itself.

\begin{figure}[t]
    \centering
    \includegraphics[width=\columnwidth]{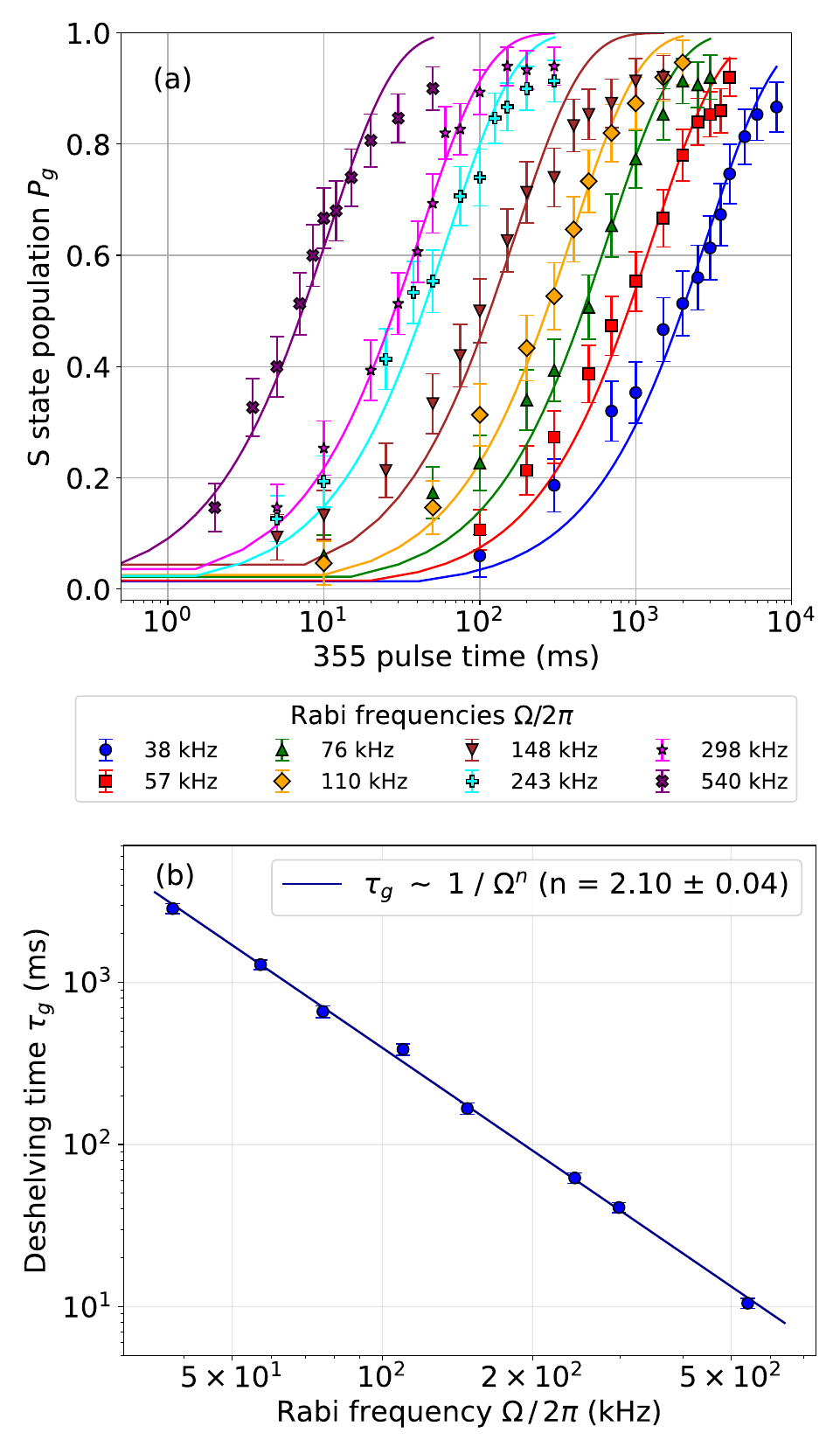}
    \caption{(a) Ions shelved in the $^2F_{7/2}$ state are deshelved in the presence of 355 nm light. For various 355 nm beam intensities (as measured through their two-photon Rabi frequencies $\Omega$), the deshelving process is well explained by an inverse exponential function. (b) The measured deshelving time monotonically decreases with 355 nm beam intensity, exhibiting a $\sim1/\Omega^2$ dependence.}
    \label{fig:deshelvingrate}
\end{figure}

To this end, we measure the $^2F_{7/2}\rightarrow ^2$\emph{S}$_{1/2}$ deshelving rate in the presence of our 355 nm Raman beams. As above, we begin by focusing each beam to a $30~\mu$m waist with 80 mW of applied power, yielding a two-photon Rabi frequency $\Omega = 2\pi\times76~$kHz. After shelving an ion in the metastable $^2F_{7/2}$ state, we apply the 355 nm beams for varying times and detect the population fraction returned to $^2S_{1/2}$. We observe that the population $P_g$ returned to the ground state qubit manifold increases with the exposure time and is well described by the inverse exponential function $P_g(t)=1 - e^{-t/\tau_g}$. Under these experimental conditions (green triangles in Fig.~\ref{fig:deshelvingrate}(a)), we measure a characteristic deshelving time of $\tau_g\approx660~$ms: considerably shorter than the natural $^2F_{7/2}$ lifetime, but still orders of magnitude longer than typical qubit-qubit interaction times.

In Fig.~\ref{fig:deshelvingrate}(a), we further extend this study to include larger and smaller Raman beam intensities. In each case, we measure the two-photon Rabi frequency $\Omega$ that results from raising or lowering the power in each beam, while keeping the beam waists fixed. We then observe the return of shelved ions to the $^2S_{1/2}$ state after the Raman beams are applied for variable times. As above, all data are well described by the inverse exponential function $P_g(t)$, with characteristic deshelving times spanning from 2 seconds (lowest-power beams) to less than 10 ms (highest-power beams).

We plot the measured deshelving time as a function of two-photon Rabi frequency in Fig.~\ref{fig:deshelvingrate}(b). The data shows an inverse quadratic dependence with laser intensity, in contrast to the inverse linear dependence expected from simple spontaneous Raman scattering \cite{ozeri2007errors}. We thus propose that our observed deshelving rates are attributable to multi-photon effects. One possible pathway is a two-photon process, scattering through the $(7/2, 1)_{5/2}$ state to a secondary level, then decaying to the ground state manifold. A more subtle possibility is the generation of large differential four-photon Stark shifts by the pulsed 355 nm laser \cite{lee2016engineering,jung2025ion}, which may bring levels in the $^2F_{7/2}$ manifold closer to resonance with the $(7/2, 1)_{5/2}$ transition. However, the precise mechanism responsible for the observed quadratic dependence is thus far unknown.

Beyond its effects on quantum simulation experiments, the deshelving induced by 355 nm Raman beams has broader implications for quantum computing architectures based on omg- or dual-type qubits in $^{171}$Yb$^{+}$ \cite{allcock2021omg,yang2022realizing}. In both schemes, it is presumed that qubits stored in metastable states are unaffected by gate operations applied to ground state qubits, due to the large mismatch in qubit frequencies. Although this has been experimentally checked for short times in the metastable $^2F_{7/2}$ manifold \cite{yang2022realizing}, our measurements indicate that 355 nm ground-state operations will directly lead to coherence loss in metastable qubits at longer times. Fortunately, given the long timescales and potential multi-photon processes involved, it is likely that this mechanism is specific to $F$-state qubits in $^{171}$Yb$^{+}$ driven by 355 nm laser light rather than a general concern for omg- or dual-type architectures.

\section{Summary and Outlook}
In this work, we have demonstrated metastable state shelving as a viable approach for in-situ rewiring of trapped-ion spin-spin interactions. Following shelving, the resulting interaction pattern is consistent with removal of unwanted ions, which remain outside the qubit subspace for the full duration of quantum dynamics. While this demonstration was performed with a global shelving beam that provides probabilistic access to different shelving configurations, individually addressed shelving would enable the deterministic removal of selected qubits and the realization of spin lattice geometries such as those in Fig.~\ref{fig:concept}.

In addition to the demonstrated rewiring of interaction graphs, our approach towards metastable state shelving enables new classes of quantum simulation experiments. For instance, shelving of ions outside the qubit subspace would allow future studies of quantum evolution in systems with lattice disorder, simulations of qubit erasure processes, and the dynamics of open quantum systems with controllable couplings to the environment. More generally, our method provides a framework to harness metastable states as a resource for scalable quantum simulations of complex spin lattice Hamiltonians.

Finally, our study has found that shelved ions remain in the metastable state for many orders of magnitude longer than the coupling timescale between ions, under typical experimental conditions. This establishes the maximum available timescale for using metastable state shelving when driving quantum dynamics. Further, we find that the metastable state decay rate scales quadratically with the laser power applied during a quantum simulation. Since the spin-spin interaction strength $J_{ij}$ likewise scales quadratically with applied laser power, this orders-of-magnitude separation of timescales remains valid over the range of accessible laser powers in our experiment.

Although high-power Raman driving rarely disrupts the shelved lattice configuration for small numbers of ions, off-resonant deshelving effects will become more significant as the system size increases. We expect the probability of maintaining $m$ ions in a shelved state to decay as $e^{-mt/\tau_g}$, where $\tau_g$ is the deshelving time. For instance, if we generate a Kagom\'e lattice by shelving $m=25$ sites in an $N=100$ ion crystal and apply our measured $\tau_g \approx 660$~ms, this corresponds to a $1/e$ lattice stability time of $25~$ms. This timescale remains notably longer than the typical decoherence time of \mbox{$\sim 5$~ ms} in trapped-ion analog quantum simulators, indicating that off-resonant deshelving is not expected to be a limiting factor at the 100-ion level.

When scaling to very large quantum systems, it may be desirable to extend the metastable state lifetime even further. We propose that the precise mechanism which deshelves ions from the $^2F_{7/2}$ state must first be identified. If this process involves off-resonant excitation from the 355 nm Raman beam followed by subsequent decays (as suspected), one approach would be further laser addressing to close these transitions and repump ions back into the $^2F_{7/2}$ state. Alternatively, one might consider different wavelengths for driving spin-spin interactions which are further detuned from transitions accessible to ions shelved in $^2F_{7/2}$. With such mitigation strategies in place, metastable state shelving would enable a straightforward and reconfigurable approach towards quantum simulation in large trapped-ion arrays.

\begin{acknowledgments}
We are grateful to Patrick McMillin for early discussions and to Wes Campbell for discussions related to metastable state deshelving. This work was supported by the Gordon and Betty Moore Foundation, grant DOI 10.37807/GBMF12963 and by the National Science Foundation under Grant No. PHY-2412878.
\end{acknowledgments}

\section*{Data Availability Statement}
The data that support the findings of this article are publicly available \cite{jung2026data}.

\begin{appendix}
\renewcommand\thefigure{\thesection.\arabic{figure}}    

\section{$^2S_{1/2}\rightarrow ^2$$D_{5/2}$ Spectroscopy}
\label{sec:appx}
\setcounter{figure}{0}    

\begin{figure}[h!]
    \centering
    \includegraphics[width=\columnwidth]{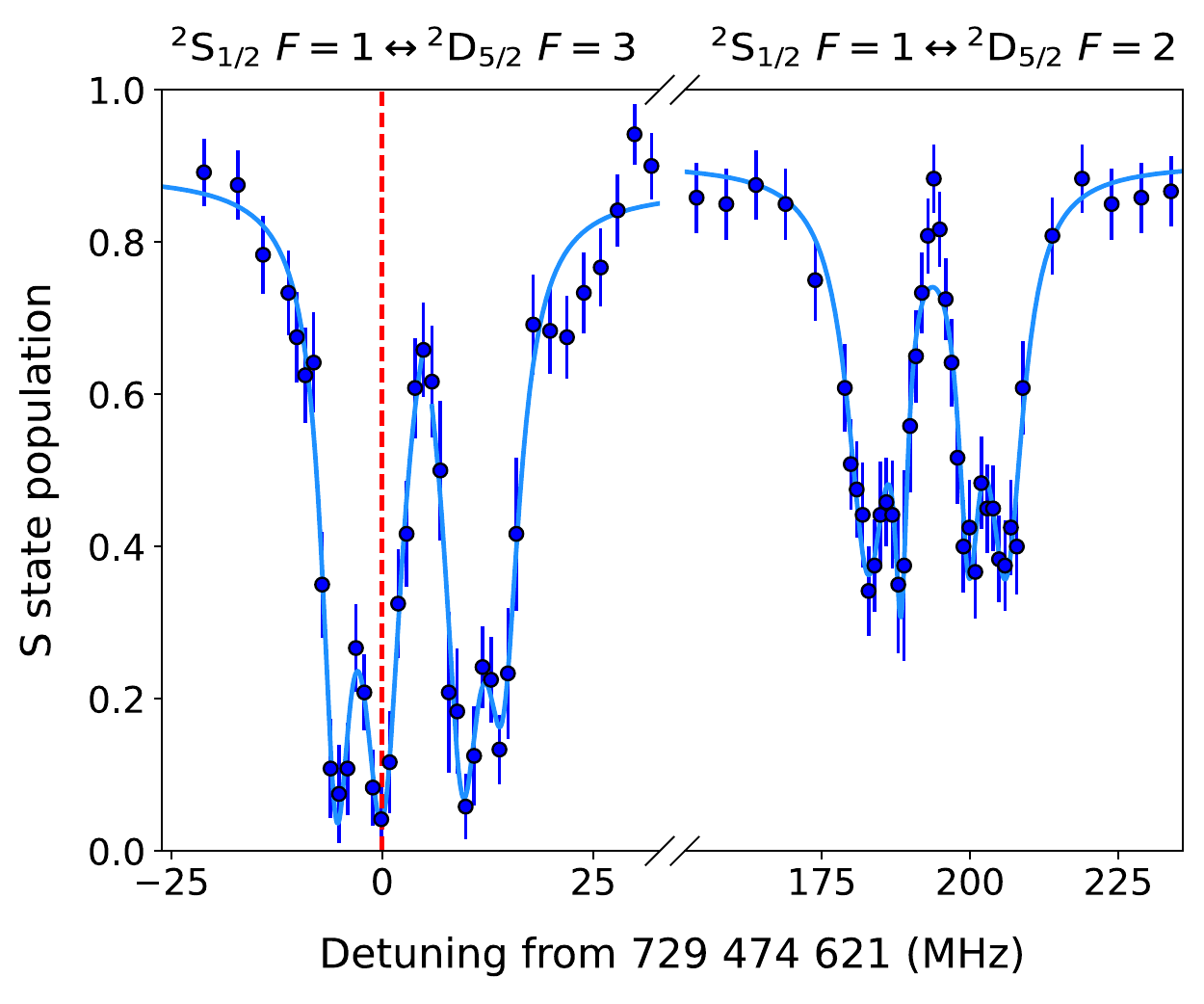}
    \caption{Spectroscopy of the 411 nm $^2S_{1/2}\rightarrow^2D_{5/2}$ transitions in a background magnetic field of 3.6 G. We optically pump to the $^2F_{7/2}$ state by driving our strongest observed resonance, corresponding to $\Delta m_F=-1$ transitions between the $^2S_{1/2}~\ket{F=1}$ and $^2D_{5/2}~\ket{F=3}$ manifolds (red dashed line).}
     \label{fig:spectroscopy}
\end{figure}

We perform spectroscopy to characterize the transition frequencies between $^2S_{1/2}\,\ket{F = 1}$ and $^2D_{5/2}\,\ket{F = 2 \text{ or } F = 3}$, including the Zeeman sublevels, in the presence of our 3.6 G magnetic field. Figure~\ref{fig:spectroscopy} shows all accessible transitions in both the $^2D_{5/2}\,\ket{F = 2}$ and $^2D_{5/2}\,\ket{F = 3}$ states, which we measure to have a hyperfine splitting of approximately $2\pi \times 190$~MHz. Each data point is averaged over 120 repetitions and includes statistical uncertainties due to $\approx$1 MHz laser frequency fluctuations and 5\% laser intensity noise. Due to significant power broadening, we observe a transition linewidth of approximately $2\pi \times$ 5 MHz, much larger than the predicted natural linewidth of $2\pi \times$ 22 Hz. For the experiments reported in this work, we shelve using the strongest measured transition between the $^2S_{1/2}$ and $^2D_{5/2}$ states (red dashed line in Fig.~\ref{fig:spectroscopy}).

\end{appendix}

\clearpage
\bibliographystyle{prsty}
\providecommand{\noopsort}[1]{}\providecommand{\singleletter}[1]{#1}%


\end{document}